\documentclass{aastex63}

\shorttitle{CMB Distortion}
\shortauthors{Corian\`o and Frampton}
\graphicspath{{./}{figures/}}

\begin{document}

\title{ Does CMB Distortion Disfavour Intermediate Mass Dark Matter?}
\correspondingauthor{Paul H. Frampton}
\email{paul.h.frampton@gmail.com}
\author{C. Corian\`o}
\affiliation{Department of Mathematics and Physics {\it ``E. De Giorgi''} , University of Salento, Via per Arnesano, CP-I93, I-73100, Lecce, Italy}
\affiliation{INFN, Sezione di Lecce, Via per Arnesano, CP-193, I-73100, Lecce, Italy}

\author{P.H. Frampton}
\affiliation{Department of Mathematics and Physics {\it ``E. De Giorgi''} , University of Salento, Via per Arnesano, CP-I93, I-73100, Lecce, Italy}
\affiliation{INFN, Sezione di Lecce, Via per Arnesano, CP-193, I-73100, Lecce, Italy}

\begin{abstract}
\noindent
We investigate the published constraints on MACHOs
in the mass region $10^2 - 10^5 M_{\odot}$ and their possible contribution
to dark matter.  We focus on constraints which rely on the accretion
of matter which emits X-rays that can lead, after 
downgrading to microwaves, to distortion of the CMB spectrum and isotropy.
The most questionable step in this chain of arguments is the use of
overly
simplified accretion models.
We compare how the same accretion models apply to X-ray observations
from supermassive black holes SMBHs, M87 and Sgr A*.
The comparison of these two SMBHs with intermediate mass MACHOs suggests that the latter
could, after all, provide the constituents of all the dark matter. We discuss the status
of other constraints on IM-MACHOs.

\end{abstract}


\section{Introduction} 

\noindent
Our ignorance of the nature of dark matter has led to theoretical suggestions of
candidates for dark matter constituents which famously range in mass over one
hundred orders of magnitude. These candidates can usefully be divided into
two classes: those which can be discovered by terrestrial experiments and
those which cannot. In the first class are new elementary particles for which
huge resources have been invested in so-far unsuccessful attempts at their
detection.

\bigskip

\noindent
An astrophysical candidate proposed \citep{chapline}, \citep{framp2010}, \citep{framp2015},\citep{chapline2}
for the consitituent of the cosmological dark matter is the Primordial Intermediate Mass Black Hole (PIMBH)
with mass in the range $10^2-10^5 M_{\odot}$. The best observational method positively to detect PIMBHs
appears to be by microlensing experiments pointing towards the Magellanic Clouds, following
the earlier success of finding lighter mass examples \citep{alcock}.

\bigskip

\noindent
One deep reason to believe that dark matter is astrophysical is that, if we regard the
visible universe as an isolated thermodynamic system, its entropy is expected to
grow according to the second law of thermodymamics. Although the question of whether
or not it is "isolated" may be debated, we believe it is reasonable to assume that
the complex dynamics of the early universe obey this statistical law. In comparison, 
no microscopic dark matter candidate such as a new elementary particle
can make a fractionally significant contribution to the entropy of the universe.

\bigskip

\section{CMB Distortion}

\noindent
The hypothesis made in \citep{framp2015}, \citep{chapline2} is that
$f_{DM}\sim100\%$ of the dark matter in the Milky Way halo is made up of PIMBHs
in the mass range $10^2-10^5M_{\odot}$. Examining the literature
one finds that this hypothesis appears to be strongly disfavoured by the
use of Bondi-like accretion models\cite{bondi},\citep{bondi2}
which lead to limits which have been placed \citep{ricotti}, \citep{ali}, \citep{carr}
on the basis of CMB distortion. Are these limits correct and if not, why not?

\bigskip

\noindent
The CMB spectrum was first measured with high accuracy by the Far Infrared Absolute 
Spectrophotometer (FIRAS) mounted on the Cosmic Background Explorer (COBE) satellite
\citep{fixsen}. The result is an extremely accurate black-body spectrum, the  most
accurate such spectrum ever measured. Also the isotropy of the CMB is measured
accurately and found to be uniform to the high level of 1 part in $10^5$, at which point important 
tiny homogeneities appear which are interpreted as the seeds of structure
formation. Any theoretical model is tightly constrained
by agreement with the CMB spectrum and isotropy.

\bigskip

\noindent
Calculation of the limits on MACHOS from lack of CMB distortion 
was pioneered in \citep{ricotti} and makes four steps (1) Calculate the
amount of material accreted on to the PIMBHs; (2) Calculate the amount of X-rays
emitted by this accreted matter; (3) Calculate the downgrading of these X-rays to
microwaves due to Thomson scattering and cosmic expansion; (4) Investigate
how much these microwaves distort the CMB spectrum and isotopy. Later in the present article, 
of these four steps we shall mainly query the validity of step (1).

\bigskip

\noindent
Starting with the paper\citep{ricotti}, from its Figure 9 (left) we deduce the
limit for the mass range $10^2-10^5M_{\odot}$ that $f_{DM} < 0.01\%$ which strongly disfavours the prediction $f_{DM}\sim 100\%$. Bondi-like quasi-spherical accretion was more recently used again in \citep{ali},
together with a reduction by a couple of orders of magnitude in the radiation efficiency, as an attempt
to correct some of the deficiencies in \citep{ricotti}. However, simliarly to \citep{ricotti}, we deduce
from Figure 14 in \cite{ali} that for $M>10^3M_{\odot}$ there is a
limit $F_{DM}<0.01\%$, again strongly disfavouring $f_{DM}=100\%$.

\bigskip

\noindent
The limits provided in \citep{ricotti}, \citep{ali} and other papers are frequently cited
in reviews, see {\it e.g.} Figure 10 in \citep{carr}, and such an exclusion plot is
almost invariably included as one slide in 
any review talk at a dark matter conference. As a result,  the widespread belief is that ($\sim 100\%$)DM = PIMBHs
is ruled out. We shall now proceed seriously to query whether such a conclusion is
warranted.

\bigskip

\section{Observations of M87 and Sgr A*}

\bigskip

\noindent
 There are a few reasons to question whether the environments of SMBHs such as M87
 and Sgr A* are sufficiently similar to those of dark matter PIMBHs
 that their direct comparison is sensible.

\noindent
(i) Sgr A* and M87 may have formerly been AGNs (Active Galactic Nuclei) \citep{macroni} when they
would likely have radiated more than at present. Such AGN radiation may not have been radially
symmetric\citep{russell}. The SMBHs have feedback effects which create a duty cycle. For AGNs there
is evidence that this duty cycle depends inversely with mass\citep{shankar}.\\
(ii) If the SMBHs were formed by mergers of smaller black holes, this could have disrupted their
gas environment, although SMBHs could instead be primordial.
(iii) The CMB distortion is believed to occur at redsifts $100 < Z < 500$ \citep{ali}, \citep{horowitz}
when the universe was denser, facilitating accretion.\\
(iv) SMBHs have angular momentum, while for PIMBHs the question of their spin 
is model dependent but in some models \citep{deluca} is expected to be close to zero.

\bigskip

\noindent
In this section we shall make the assumption, {\it faute de mieux}, that the environments of SMBHs and
PIMBHs, while certainly somewhat different, are sufficiently similar that their comparison is meaningful.

\bigskip

\noindent
Observations of X-rays from supermassive black holes (SMBHs) can shed useful
light on whether Bondi-like accretion 
on to a PIMBH is
a sensible model. By Bondi-like, we include all models that are
spherically symmetric, quasi-spherical or involve radial inflow.

\bigskip

\noindent
The X-ray observations have included two supermassive black holes (SMBHs), 
namely  M87 with mass
$M\simeq 6.5 \times 10^9 M_{\odot}$ and Sgr A* with mass $M\simeq 4.1 \times10^6M_{\odot}$. 
The reason these two black holes are the
ones most thoroughly studied observationally in X-rays is simply that they are the two biggest
black holes on the sky. M87, the 87th in the Messier catalogue of objects and discovered by
Messier himself in 1781, is at a distance $16.4\times 10^6$pc and subtends 
38$\mu$as ($\mu$as = microarcsecond).
Sgr A* in the Milky Way was discovered more recently 
at $8.2\times10^3$pc and subtends 52$\mu$as.

\bigskip

\noindent
We shall show that Bondi-like
models fail badly for both these supermassive black holes
by vastly overestimating, by four orders
of magnitude, the rate of accretion.
Although these two examples
are more massive than PIMBHs ranging up to $10^5 M_{\odot}$,
we are not aware of any reason that if an accretion model fails badly
for SMBHs it should start working for PIMBHs.

\bigskip

\noindent
The accretion rate for M87, calculated from a Bondi-like model is
$\dot{M} = 0.2 M_{\odot}/y$ \citep{kuo}. X-ray observations reported in 2014,
however, correspond to a far smaller accretion rate $< 10^{-5}M_{\odot}/y$
\citep{kuo}. A similar conclusion was confirmed two years
later in 2016 in both \citep{feng} and \citep{li}. We conclude 
that the accretion rate on to M87 is four orders of magnitude
smaller than that predicted in a Bondi-type model of accretion.

\bigskip

\noindent
The SMBH Sgr A* in the Milky Way is an exceptionally faint SMBH, the faintest SMBH known
and visible only due to its proximity. In any other galaxy, Sgr A* would be invisible.
In the case of Sgr A*, already in 2000 Quateart and Gruzinov\citep{quateart}
used a Bondi-type model to predict an accretion rate $\sim10^{-4} M_{\odot}$/y but provided observational
evidence that the actual accretion rate $\sim 10^{-8}M_{\odot}$/y, like M87 showing that
the Bondi-type model overestimates the accretion rate by four orders of magnitude,
This observational accretion rate on to Sgr A* was confirmed in 2018\citep{bower}.

\bigskip

\noindent
Let us next revisit the PIMBH limits derived from CMB distortion, with now the new assumption
Bondi-type models overestimate accretion rates by four orders of magnitude. It is straightforward to revise
the previous limits based on CMB distortion by replacing the published $f_{DM}$ by $f_{DM}^{(revised)} = 10^4f_{DM}$.

\bigskip

\noindent
In Figure 9(left) of \citep{ricotti} the revised $f_{DM}^{(revised)}$ is consistent
with $f_{DM}^{(revised)} \simeq 100\%$ and with a private
communication \citep{ostriker} from the senior author of \citep{ricotti}.
In Figure 14 of \citep{ali} the revised $f_{DM}^{(revised)}$ is also consistent
with $f_{DM}^{(revised} \simeq 100\%$.

\bigskip

\noindent
It is beyond the scope of this article to provide a better model for accretion than
the Bondi-type models. It seems probable that the accretion on to PIMBHs is not
approximately quasi-spherical or with radial inflow. 

\bigskip

\section{Wide Binaries}

\noindent
There exist in the Milky Way pairs of stars which are gravitationally bound binaries with a separation more than 
0.1pc, even more than 1.0pc. These wide binaries retain their original orbital parameters unless compelled to change them by 
gravitational influences, for example, due to nearby PIMBHs.

\bigskip

\noindent
Because of their very low binding energy, wide binaries are particularly sensitive to gravitational perturbations 
and can be used to place an upper limit on, or to detect, PIMBHs. The history of employing this ingenious 
technique is regretfully checkered. In 2004 a fatally strong constraint was claimed by an Ohio State University 
group \citep{yoo} in a paper entitled ``The End of the MACHO Era".

\bigskip

\noindent
Five years later in 2009, however, another group this time from Cambridge University \citep{quinn}
re-analysed the available data on wide binaries and reached a quite different conclusion. They questioned 
whether any rigorous constraint on MACHOs could yet be claimed, especially as one of the most important 
binaries in the earlier sample had been misidentified.

\bigskip

\noindent
Confirming wide binaries is challenging, as they are quite rare and hard to distinguish from
merely chance associations. Obtaining accurate orbital parameters is also difficult when only
a small fraction of one orbit can be observed.

\bigskip

\noindent
Only three studies of wide binaries have yet appeared in the literature, the two in 2004 and 2009
and one more recent in 2014. This further study of wide binaries \citep{monroy} also attempted to 
place limits on MACHOs, starting from a longer list of $\sim 100$ candidate binaries. Upon closer
inspection, the majority were called into question. It is mildly surprising that in the last seven
years, no fourth independent analysis has appeared.

\bigskip

\noindent
Unlike microlensing which has positive signals\citep{alcock}, wide binary analysis is a null experiment
where no MACHO-binary interaction has been recorded and this fact tends to physicists remaining skeptical 
of any limits claimed. Concerning the third and still most recent 2014 analysis\citep{monroy}, one of the authors of of the second 2009 analysis \citep{quinn} has
remarked \citep{belakurov} that he will remain skeptical until their orbits are re-calculated using
Gaia data.

\bigskip

\noindent
In conclusion about wide binaries, these constraints may possibly be shown on exclusion plots 
but only with the warning that they are not accepted as robust because of the several
uncertainties.

\bigskip

\section{X-ray and radio sources}. 

\noindent
In one stand-alone paper \citep{gaggero}, a study was made of the X-ray and
radio emission from the Galactic Ridge region of the Milky Way within 2 kpc of the galactic center
which, if $10 M_{\odot}$ PBHs make up all the dark matter. should contain $\sim 10^9$ PBHs.

\bigskip

\noindent
This paper claims that PBHs cannot make up all the dark matter, by a statisical
discrepancy of $5\sigma$.  However, they use a Bondi-type model
of accretion in their Eq.(1) with a multiplicative factor $\lambda = 0.02$. If, as suggested in our text,
this is changed to $\lambda=10^{-4}$ the bounds are relaxed by a factor 200 whereupon study
of their Fig. 1 reveals that the PBHs could then comprise all the dark matter without any such
discrepancy. 

\bigskip

\noindent
The same paper \citep{gaggero} considers radio emission from an assumed jet
associated with each of the black holes. They arrive then at a staggering $40\sigma$
discrepancy in $1.4 GHz$ radio waves compared to the VLA catalog. However, this
huge discrepancy arises only because of an assumption about jets and could be
regarded as much an argument against such jets as against PBH dark matter.

\bigskip

\section{Supernova microlensing}.  

\noindent
In \citep{zuma}, the absence of microlensing which could provide magnification
of SNe 1A supernovas was converted into an upper bound on MACHOs . These authors claimed that only
$< 35\%$ of the total matter content could be MACHOs, compared to the required dark matter which is $84\%$.

\bigskip

\noindent
In their analysis, they focus on a parameter $\alpha$ defined by $\alpha = \Omega_{PBH}/\Omega_m$
and fix $\Omega_m$ by using a value extracted from CMB and BAO measurements. They require a lens
size less than the Einstein radius which implies that $M_{PBH} > 0.01 M_{\odot}$.

\bigskip

\noindent
Two samples of SNe data are examined to arrive at independent values for $\alpha$ which would need
to be $\alpha \sim 0.84$ for the case that PBHs constitute all the dark matter. One sample is from
the Joint Likelihood Analysis (JLA) in reference \citep{betoule}, the other is from Union data in
reference \citep{suzuki}.

\bigskip

\noindent
In \cite{zuma}, the constraints are found to be, at 95\% confidence level, that $\alpha<0.352$ and
$\alpha<0.372$ respectively for JLA and Union. These translate into exclusion of all dark matter
being MACHOs with $>0.01M_{\odot}$ by $4.79\sigma$ for JLA and by $4.54\sigma$ for Union.

\bigskip

\noindent
However, this claim in \citep{zuma} has been subsequently disputed in \citep{garcia} for several
reasons, as follows. 

\bigskip

\noindent
Firstly, in \citep{zuma} the value of $\Omega_m$ was held fixed while $\alpha$ was varied whereas
the two are highly correlated. For example, the value of $\Omega_m$ set by the supernova
data alone is quite different so the chosen prior on $\Omega_m$ strongly overconstrained
the result for $\alpha$.

\bigskip

\noindent
Secondly, the characteristic size for the supernova is bigger than assumed in \citep{zuma}
and when this is taken into account the constraint is weakened by an order of magnitude.

\bigskip

\noindent
Thirdly and lastly, using a more realistic broad lognormal mass distribution rather than
a monochromatic one, the constraints become even further diluted.

\bigskip

\noindent
Taking all these considerations into account, the authors of \citep{garcia} conclude that
the supernova data are consistent with 100\% of the dark matter being made from PBHs.

\bigskip

\noindent
In conclusion about supernova microlensing, these constraints should be shown on exclusion plots 
only with a warning that they are not yet universally accepted.

\bigskip

\section{Ultra Faint Dwarf Galaxies (UFDGs)}. 

\noindent
UFDGs have a high ratio of total dynamical mass to stellar mass so that they contain an unusually
high fraction of dark matter. An extreme case is Eridanus II with 99.7\% dark matter.
If dark matter is 100\% PBHs, heavier on average than the
stars, then two-body collisions between PBHs and stars will give kinetic energy to
the stars and heat them up, tending to increase the size of the galaxies. Observations
of UFDGs can be used to put limits on such theories of dark matter.

\bigskip

\noindent
In the paper \citep{steg} a study of UFDGs
is made and places strong constraints on the range of masses of PIMBHs
which are allowed to comprise all the dark matter.A sample of 17 UFDGs
is used, and  various versions of mass fucnction {\it e.g.} monochromatic
or lognormal.

\bigskip

\noindent
From the survival of the stellar cluster Eridanis II, and the entire stellar populations
of UFDGs, no solution is found for any mass function of PBHs. In all cases, the
PBHs warm up the stellar systems to be too large to agree with observations.

\bigskip

\noindent
There is a number of possible shortcomings of the detailed analysis in \citep{steg}
which should be borne in mind:

\bigskip

\noindent
(i).  The stellar observables in the lightest UFDGs have low statistics.

\bigskip

\noindent
(ii).  Some UFDGs may be baryon dominated rather than dark matter dominated.

\bigskip

\noindent
(iii).  In \citep{steg}, only the case DM = 100\% PBHs is studied. If the DM has
a significant collisionless component, the analysis is invalid.

\bigskip

\noindent
(iv).  The r\^ole of gas, or a possible central IMBH, is ignored.

\bigskip

\noindent
(v).  It was assumed that UFDGs are spherical and in dynamical equilibrium.

\bigskip

\noindent
Finally, the analysis in \citep{steg} discusses the range $(1 - 100) M_{\odot}$ for
the PBHs. Intermediate mass PBHs in the range $(100 - 100,000) M_{\odot}$ are
not expected to play other than the central role in UFDGs, for stability reasons
analogous to the disk stability\citep{xu} in full-sized galaxies.

\bigskip

\section{Conclusions}

\noindent
If we take into account all proposed constraints on the allowed
fraction $f_{DM}$ of dark matter which is allowed as a function
of mass in the intermediate mass range 
$100M_{\odot} \leq M_{PBH} \leq 10^5 M_{\odot}$, and
discard all constraints which are in any way questionable
or model-dependent, our considered opinion is that $0 \leq f_{DM} \leq 1$
is allowed throughout this entire mass range. See Figure 1.
We do accept that, within
the Milky Way, PBHs with
$M_{PBH} > 10^6 M_{\odot}$ are excluded
by the consideration of disk stability \citep{xu}.

\begin{figure}[t]
\centering
\includegraphics[scale=0.3]{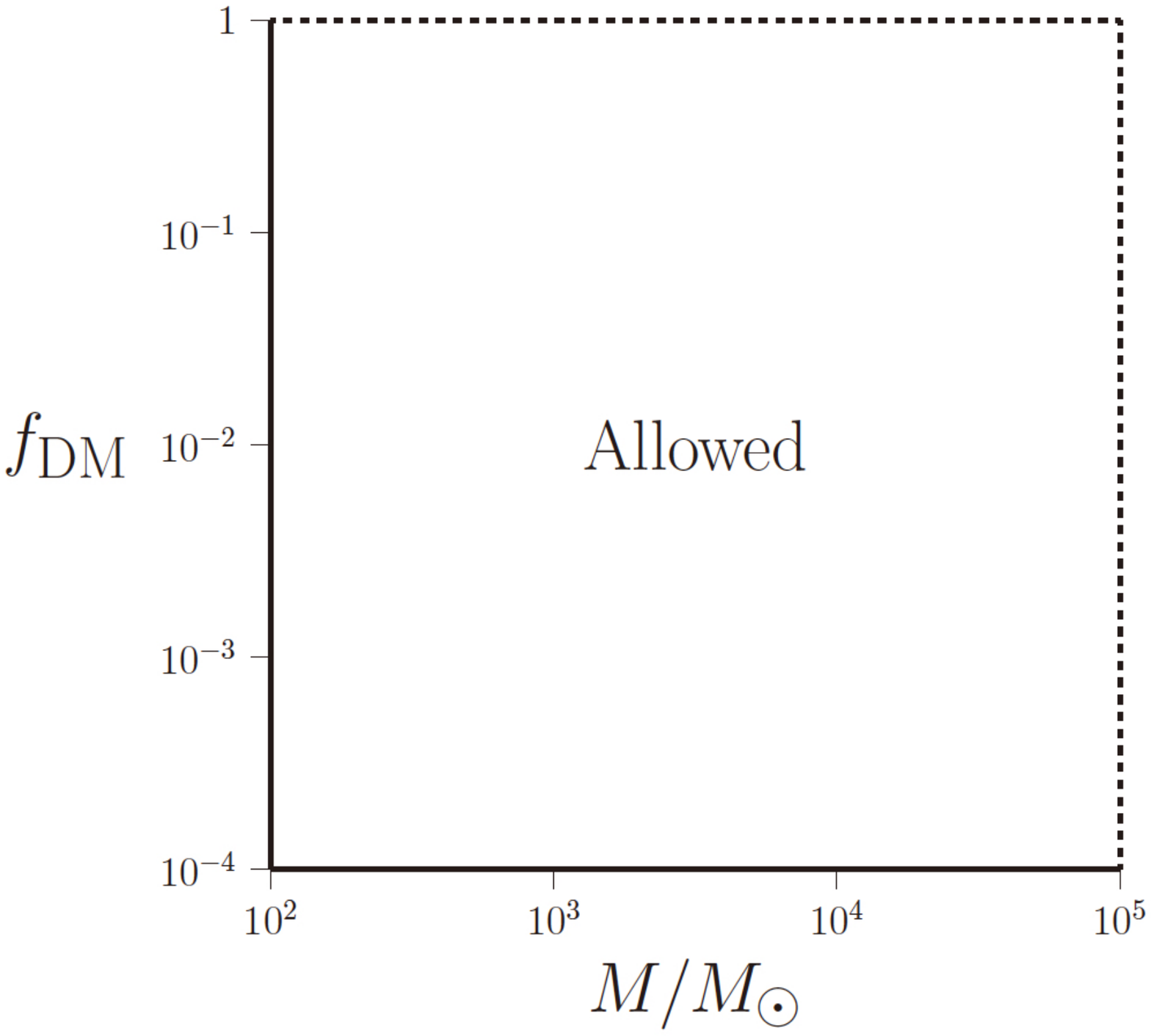}
\caption{Allowed region for $f_{DM}$ versus
$M_{PIMBH}$ with only ironclad constraints included. This may be compared and contrasted with Fig. 10 in \citep{carr}}
\end{figure}

\bigskip

\noindent
As we have discussed, a handful of well-cited papers in the astronomy 
literature imply that CMB
distortion very strongly constrains intermediate mass dark matter, so their
answer to our title question is a categoric yes. 

\bigskip

\noindent
However, by making a comparison with data from
X-ray observations of two supermassive black holes, and making
the strong but defensible assumption that their accretion dynamics
are sufficiently similar to those of intermediate mass, we have
cast doubt on this conclusion. 
Hence we believe the answer to our title question may be no,
and that intermediate mass black holes can make a significant
contribution to the dark matter.

\bigskip

\noindent
For completeness we have discussed also published constraints arising
from (i) wide binaries; (ii) X-ray and radio sources; (iii) superniva microlensing;
(iv) ultra-faint dwarf galaxies. Although excellent pieces of work, all of which merit
further study, in their present form we do not feel it necessary to change our
opinion about the freedom still allowed by all of the present data.


\bigskip



\begin{thebibliography}{99}
{
\bibitem[\protect \citeauthoryear{Alcock et al}{2000}]{alcock}
Alcock, C, et al. (MACHO Collaboration), 2000,
Astrophys.J. {\bf 542,}  281}

\bibitem[\protect \citeauthoryear{Ali-Ha\"{i}moud and Kamionkowski}{2017}]{ali}
Ali-Ha\"{i}moud, Y and Kamionkowski, M. , 2017,
Phys. Rev. D95, 043534 

\bibitem[\protect \citeauthoryear{Belakurov}{2021}]{belakurov} 
Belakurov, V., private communication, January 2021

\bibitem[\protect \citeauthoryear{Betoule et al}{2014}]{betoule}
Betoule, M., et al., (SDSS Collaboration), 2014,
Astron.Astrophys. 568, A22 

\bibitem[\protect \citeauthoryear{Bondi and Hoyle}{1944}]{bondi}
Bondi, H. and Hoyle, F., 1944, MNRAS 104, 273 

\bibitem[\protect \citeauthoryear{Bondi}{1952}]{bondi2}
Bondi, H., 1952, MNRAS  112, 195 

\bibitem[\protect \citeauthoryear{Bower et al}{2018}]{bower}
Bower, G.C., et al., 2018,
Astrophys.J. 868, 101

\bibitem[\protect \citeauthoryear{Carr et al}{2020}]{carr}
Carr, B., K.Kohri, K, Y. Sendouda, Y and Yokohama,J, 2020,
arXiv:2002.12778[astro-ph.CO]

{
\bibitem[\protect \citeauthoryear{Chapline}{1975}]{chapline}
Chapline, G.F., 1975, Nature 253, 251}

\bibitem[\protect \citeauthoryear{Chapline and Frampton}{2016}]{chapline2}
Chapline, G.F. and Frampton, P.H., 2016, JCAP  11:042 

\bibitem[\protect \citeauthoryear{De Luca et al}{ 2019}]{deluca}
De Luca, V.,  Desjacques, V., Franciolini, G., Malhotra, A. and Riotto, A., 2019,
JCAP 05:018

\bibitem[\protect \citeauthoryear{Feng, Wu and Lu}{2016}]{feng}
Feng, J.,  Wu Q., and Lu, R.-S., 2016,
Astrophys. J. 830, 6

\bibitem[\protect \citeauthoryear{Fixsen et al}{1996}]{fixsen}
Fixsen, D.J., et al.,
Astrophys. J. 473, 576 

\bibitem[\protect \citeauthoryear{Frampton et al}{2010}]{framp2010}
Frampton, P.H., Kawasaki M., Takahashi, F and Yanagida, T. 2010,
JCAP 04:023 

\bibitem[\protect \citeauthoryear{Frampton}{2015}]{framp2015}
Frampton, P.H., 2016, Mod. Phys. Lett. A31, 1650093 

\bibitem[\protect \citeauthoryear{Gaggero et al}{2017}]{gaggero}
Gaggero, D.,Bertone, G., Calore, F.,  Connors, R.M.T.,  Lovell, M.  et al., 2017,
Phys.Rev.Lett. 118, 241101

\bibitem[\protect \citeauthoryear{Garcia-Bellido, Clesse and Fleury}{2018}]{garcia}
Garcia-Bellido, J., Clesse, S.  and Fleury, P., 2018,
Phys.Dark Univ. 20, 95 

\bibitem[\protect \citeauthoryear{Horowitz}{2016}]{horowitz}
Horowitz, B., 2016,
arXiv:1612.07264 [astro-ph.CO]

\bibitem[\protect \citeauthoryear{Kuo et al}{2014}]{kuo}
Kuo, C.Y., et al., 2014,
Astrophys.J.Lett.  783, L33 

\bibitem[\protect \citeauthoryear{Li, Yuan and Xie}{2016}]{li}
Li,  Y.-P.,  Yuan, F and Xie, F-G, 2016,
Astroph. J. 830, 78 

\bibitem[\protect \citeauthoryear{Macroni et al}{2004}]{macroni}
Macroni,, A., Risaliti, G., Gilli, R., Hunt, L.K., Maiolino, R., et al., 2004,
MNRAS  351, 169 

\bibitem[\protect \citeauthoryear{Monroy-Rodriguez and Allen}{2014}]{monroy}
Monroy-Rodriguez, M.A. and Allen, C., 2014,
Astrophys.J. 790, 159 

\bibitem[\protect \citeauthoryear{Ostriker}{2016}]{ostriker}
Ostriker, J.P. private communication, December 2016.

\bibitem[\protect \citeauthoryear{Poulin et al}{2017}]{poulin}
Poulin, V., Serpico, P.D., Calore, F., Cleese, S., and Kohri,K., 2017,
Phys. Rev.  D96, 083524 

\bibitem[\protect \citeauthoryear{Quateart and Gruzinov}{2000}]{quateart}
Quateart, E. and Gruzinov, A., 2000,
Astrophys. J. 545, 842 

\bibitem[\protect \citeauthoryear{Quinn et al}{2009}]{quinn}
Quinn, D.P., Wilkinson, M.I., Irwin, M.J., Marshall, J., Koch, A., and Belokurov, V.,2009,
MNRAS  396, 11

\bibitem[\protect \citeauthoryear{Ricotti, Ostriker and Mack}{2008}]{ricotti}
Ricotti, M, Ostriker, J.P., and Mack, K.J., 2008,
Astrophys J. 680, 829 

\bibitem[\protect \citeauthoryear{Russell et al}{2013}]{russell}
Russell, H.R.,  McNamara, B.R., Edge, A.C. Hogan, M.T., et al., 2013,
MNRAS  432, 530

\bibitem[\protect \citeauthoryear{Shankar, Weinberg and Shen}{2010}]{shankar}
Shankar, F., Weinberg D.H. and Shen, Y., 2010,
MNRAS  406, 1959 

\bibitem[\protect \citeauthoryear{Stegmann et al}{2020}]{steg}
Stegmann, J., Capelo, P.R.,  Bortolas, E. and Mayer, L., 2020, 
MNRAS 492, 5247 

\bibitem[\protect \citeauthoryear{Suzuki et al}{2012}]{suzuki}
Suzuki, N., et al., (Supernova Cosmology Project Collaboration), 
2012, Astrophys.J. 746, 85 

\bibitem[\protect \citeauthoryear{Xu and Ostriker}{1994}]{xu}
Xu, G.H. and Ostriker, J.P., 1994,
Astrophys.J. 437, 184

\bibitem[\protect \citeauthoryear{Yoo, Chaname and Gould}{2004}]{yoo}
Yoo, J., Chaname, J., and Gould, A., 2004,
Astrophys. J. 601, 311 

\bibitem[\protect \citeauthoryear{Zumalacarregui and Seljak}{2018}]{zuma}
Zumalacarregui, M and Seljak, U., 2018
Phys.Rev.Lett. 121, 141101 

\end{thebibliography}
\end{document}